\documentstyle[11pt,aaspp4]{article}
\begin{document}
\title{Images of Bursting Sources of High-Energy Cosmic Rays.
I: Effects of Magnetic Fields}
\author{Eli Waxman and Jordi Miralda-Escud\'e}
\affil{Institute for Advanced Study, Princeton, NJ 08540}
\authoremail{ waxman@ias.edu, jordi@ias.edu }
\slugcomment{Submitted to Ap. J. ({\it Letters})}

\begin{abstract}

It has recently been shown that the highest energy cosmic rays (CRs) may
originate in the same cosmological objects producing $\gamma$-ray bursts. This
model requires the presence of intergalactic magnetic fields (IGMF) to delay
the arrival times of $\sim 10^{20}$ eV CRs by 50 years or longer relative to
the $\gamma$-rays, of an amplitude that is consistent with other observational
constraints. Sources of CRs coming from individual bursts should be resolved
with the planned ``Auger'' experiment, with as many as hundreds of CRs for the
brightest sources. We analyze here the apparent angular and energy distribution
of CRs from bright sources below the pion production threshold (in the energy
range $10^{19}{\rm eV} < E < 4\times10^{19}{\rm eV}$) expected in this model.
This observable distribution depends on the structure of the IGMF: the apparent
spectral width $\Delta E$ is small, $\Delta E/E\lesssim1\%$, if the
intergalactic field correlation length $\lambda$ is much larger than
$1{\rm Mpc}$, and large, $\Delta E/E=0.3$, in the opposite limit $\lambda\ll
1{\rm Mpc}$. The apparent angular size is also larger for smaller $\lambda$.
If the sources of CRs we predict are found, they will corroborate the bursting
model and they will provide us with a technique to investigate the structure 
of the IGMF.

\end{abstract}

\section{Introduction}

The origin of cosmic rays (CRs) with energy $E>10^{19}{\rm eV}$ is unknown.
Most of the sources of cosmic rays that have been proposed have difficulties 
in accelerating CRs up to the highest observed energies (e.g.,
\cite{huge,obj1,obj2}). Recently, a new model has been proposed where the same
astrophysical objects responsible for $\gamma$-ray bursts (GRBs) also
produce the highest energy CRs (\cite{Wa}, \cite{Vietri},
\cite{MU}). There is increasing evidence that GRBs and the highest
energy CRs are of cosmological origin (\cite{cos}, \cite{Fly1},
\cite{AGASA2}, \cite{Wb}). In this case, the observational
characteristics of GRBs impose strong constraints on the physical
conditions in the $\gamma$-ray emitting region, which make it an ideal
site for accelerating protons up to the highest observed CR energies.
In addition, the average rate (over volume and time) at which energy
is emitted as $\gamma$-rays by GRBs and in CRs above $10^{19}
{\rm eV}$ in the cosmological scenario is remarkably comparable 
(\cite{Wa},b).

  The energy loss of CRs with $E\gtrsim10^{20}{\rm eV}$, due to 
interaction with the microwave background, imply that they 
must originate at distances $< 100$ Mpc (e.g., \cite{huge}). 
To reconcile the
detection of two CRs with $E\gtrsim10^{20}{\rm eV}$ over a
$\sim5{\rm yr}$ period (\cite{Fly1},\cite{AGASA2}) 
with the expected rate of nearby GRBs
($\sim1$ per $50{\rm yr}$ in the field of view of the CR experiments out
to $100 {\rm Mpc}$; e.g., \cite{rate}), a dispersion in arrival times
$\geq50{\rm yr}$ for CRs produced in a single burst needs to be invoked.
Such dispersion may result from deflections of CR protons by the
intergalactic magnetic field (IGMF) (\cite{Wa}, \cite{Coppi}).
The deflection angle for a proton
propagating a distance $D$ in a magnetic field $B$ with
correlation length $\lambda$ is $\theta_s \simeq 0.\arcdeg025,
(D/\lambda)^{1/2}\, (\lambda/10{\rm Mpc})\, (B/10^{-11}\, {\rm G})\,
(E/10^{20}{\rm eV})^{-1} $, and the induced 
time delay is $\tau(E) \sim D \theta_s^2/4c \sim 200 {\rm yr}
(D/100{\rm Mpc})^2\, (\lambda/10{\rm Mpc})\, (B/10^{-11}\, {\rm G})^2\,
(E/10^{20}{\rm eV})^{-2}$ (where the exact dependence of $\tau$ on
$\theta_s$ depends on the scale of the magnetic field; see below).
The required magnetic field is consistent with observational limits
(\cite{Kron,Vallee}).

  In a preceding paper (\cite{MW}), we calculated
the number of CR sources with different flux at different energies in
the model of ``Cosmic Ray Bursts''.
The energy dependent delay of the CR arrival times results in individual
sources having relatively narrow observed spectra,
since at any given time only those CRs having a fixed time delay are
observed. It was shown that in the GRB model
most of the CRs with $E\gtrsim2\times10^{20}{\rm eV}$ should come only from 
a few sources, since the short time delays implied by the small distance
to the sources allow
only a few bursts to be seen at any given time. At lower
energies many more sources are observable due to both the larger
distances over which the CRs can propagate and the much longer time
delays. Therefore, while at the highest energies bursting sources should
be identified from only a small number of CRs from their coincident
directions, many more CRs need to be detected at low energies to
identify sources. Recently, the AGASA experiment reported the presence
of 3 pairs of CRs with angular separations (within each pair) $\le2.5\arcdeg$, 
consistent with the measurement error,
among a total of 36 CRs with $E\ge4\times10^{19}{\rm eV}$ (\cite{pairs}). 
The two highest energy AGASA events were in these pairs.
Given the total solid angle observed by the experiment, $\sim2\pi{\rm sr}$,
the probability to have found 3 pairs by chance is $\sim3\%$; and, given that
three pairs were found, the probability that the two highest energy CRs are
among the three pairs by chance is 2.4\%. Therefore, this observation favors
the bursting source model, although more data are needed to confirm it.

 The detailed spectral and angular distribution of CRs from an individual
source, expected in a bursting source model, depend on the structure of the 
IGMF and on the details of the energy
loss processes. This shall be examined in this Letter, confining the
discussion to sources observed at energies below the pion production
threshold, $E\lesssim4\times10^{19}{\rm eV}$.
Sources at higher energy will be treated in a subsequent 
publication.

\section{The Propagation of CRs through the Magnetic Field}

\subsection{Large Scale Fields}

Let us first consider the case where the typical displacement of the CRs is 
much smaller than the correlation length, $\lambda \gg D\theta_s(D,E)$.
In this case, all the CRs that arrive at the
observer are essentially deflected by the same magnetic field structures. 
Therefore, if the CR energy loss during propagation was perfectly continuous
and deterministic,
all CRs with a fixed observed energy would reach the observer with exactly
the same direction and time delay. At a fixed time, the source would
appear monoenergetic and point-like, with the CR energy decreasing with time
as $t^{-1/2}$, and the deflection from the true source position increasing as 
$t^{1/2}$. However, the fact that CRs suffer random
energy losses as they propagate, owing to the production of pions and $e^\pm$ 
pairs in interactions with the microwave background, implies that in reality 
the image should be blurred both in arrival direction and energy.
Due to the random energy loss,
at any distance from the observer there is some finite spread
in the energies of CRs that are observed with a given fixed energy.
Since the deflection due to the magnetic
field is everywhere inversely proportional to the CR energy, the
fractional variation of the total deflection, $\delta\theta(E,D)/
\theta_s(E,D)$, and of the time delay, $\delta t(E,D)/\tau(E,D)$, are 
similar to the characteristic fractional energy spread along the path, 
$\sigma_E(E,D)/E$.
The spread in energy of CRs observed at a given time is also
$\Delta E/E\sim \delta t/\tau\sim \sigma_E(E,D)/E$.

The brightest sources of CRs with $E\lesssim4\times10^{19}{\rm eV}$ should
be very nearby. The typical distance $D_m(E)$ to the brightest source observed
over an energy range $\Delta E$ around $E$, with $\Delta E/E\sim 1$,
is the radius of a sphere within which the average time between bursts is
equal to the characteristic time delay $\tau[E,D_m(E)]$;
i.e., $D_m$ is determined by $4\pi D_m^3\nu\tau(E,D_m)/3=1$
where $\nu$ is the burst rate per unit volume. 
Thus,
\begin{equation}
D_m(E)\simeq 30\left({\nu\tau_0\over10^{-6}{\rm Mpc}^{-3}}\right)^{-1/5}
E_{19}^{2/5}\,{\rm Mpc},
\label{Dm}
\end{equation}
where $E=10^{19}E_{\rm 19}{\rm eV}$ and $\tau_0\equiv\tau(E=10^{20}{\rm eV},
D=100{\rm Mpc})$. The sources of the events that have been detected
above $10^{20}{\rm eV}$ must lie within $100{\rm Mpc}$, requiring
$\nu\tau_0>10^{-6}{\rm Mpc}^{-3}$. From eq.\ (\ref{Dm}) we see that at lower
energies the brightest sources will be even closer.
These CRs from bright sources will suffer energy loss only by pair production,
because at $E < 5\times 10^{19}$ eV
the mean-free-path for pion production interaction
(in which the fractional energy loss is $\sim10\%$) is larger than 
$1{\rm Gpc}$. Furthermore, the energy loss due to pair production 
over $100{\rm Mpc}$ propagation is only $\sim5\%$.

 The fractional energy loss in a pair production interaction is
$\sim m_e/m_p$. The average total energy loss after $N$ interactions is
$\delta E/E \simeq N\, m_e/m_p$, and the dispersion in the energy loss
is $\sigma_E/E\simeq \sqrt{N}\, m_e/m_p = (m_e/m_p\, \delta E/E)^{1/2}$.
The spectral width and apparent angular size of bright sources at 
$E<4\times10^{19}{\rm eV}$, for which $\delta E/E\lesssim5\%$,
are therefore very small,
$\Delta E/E\sim\delta\theta/\theta_s\sim\sigma_E/E\lesssim1\%$.
On the other hand, CRs of much higher energy can lose a substantial fraction
of their energy in a single pion production,
resulting in a large dispersion in the energy loss
for a propagation distance of a few tens of megaparsecs.
This implies that the width of the energy spectrum and the
angular size of the image should be much wider at high energies.
Detailed predictions of the spectral shape due to the random energy loss
caused by pion production will be presented in a subsequent publication.

\subsection{Small Scale Fields}

Let us now consider the case where the typical displacement of the CRs is 
much larger than the correlation length, $\lambda \ll D\theta_s(D,E)$.
In this case, the deflection of different CRs arriving at the observer
are essentially independent. Even in the absence of any energy loss there 
are many paths from the source to the observer for CRs of fixed energy $E$
that are emitted from the source at an angle 
$\theta\lesssim\theta_s$ relative to the source-observer line of sight. Along
each of the paths, CRs are deflected by independent magnetic field structures.
Thus, the source angular size would be of order $\theta_s$
and the spread in arrival times would be comparable to the characteristic 
delay $\tau$, leading to $\Delta E/E\sim1$ even when there are no random
energy losses.

  To allow for an analytical treatment of the case $\theta_sD\gg\lambda$, 
we consider CRs below the pion production threshold and
neglect their energy losses due to
$e^\pm$ pair production (which is, as shown above, a good approximation
for bright sources of CRs with $E\lesssim4\times10^{19}{\rm eV}$).
Since each CR arriving at the observer follows a different path,
different CRs sample different ``realizations'' of the IGMF.
The distribution of observed deflections
and time delays for an observer at a distance $r$ from the source is therefore 
given by the probability density $P(\varphi,t;r,E)$ for a CR propagating 
to a distance $r$ to encounter a magnetic field structure that would cause
a time delay $t\equiv T-cr$ (where $T$ is the time since 
the CR was emitted) and deflection angle $\varphi$ relative to the
direction to the source. 

Let us divide $r$ in small intervals $\Delta r$ such
that $\lambda<\Delta r\ll r$. The deflections in different intervals
are independent, since $\Delta r$ is larger than the field 
correlation length.
The evolution of the probability density $P$ may then be
described as a Markov process that changes $P$ at each interval $\Delta r$.
Since the number of independent deflections, $r / \Delta r$, is very large,
the relative change in $P$ over $\Delta r$ is small, and we may approximately
describe its evolution by a differential equation
obtained by formally taking the limit $\Delta r\rightarrow0$.

In what follows we replace the variable $\varphi$ with $\zeta=\varphi^2$,
$P(\zeta,t;r,E)\equiv P(\varphi,t;r,E)d\varphi/d\zeta$.
Since $d\zeta = \varphi d\varphi/2$, $P(\zeta,t;r,E)$ is the probability per 
unit solid angle. In the absence of magnetic field, as a CR moves to a 
distance $r+\Delta r$ its angle $\varphi$ decreases by $\Delta\varphi 
= - \varphi\, \Delta r/r$, i.e., $\Delta\zeta = -2\zeta\, \Delta r/r$,
and its time delay increases by $\Delta t = \zeta\, \Delta r/2c$.
Thus, in the absence of magnetic deflection the
evolution of $P$ is determined by
\begin{equation}
{dP\over dr} = {\partial P\over \partial r} - {2\zeta\over r}
{\partial P\over \partial \zeta} + {\zeta\over2c} {\partial P \over
\partial t} = {2P\over r} \quad,
\label{dPp}
\end{equation}
where the right-hand-side is due to the change in the volume element
$d\zeta\, dt$ for CRs moving from $r$ to $r+\Delta r$. In the presence
of an IGMF, a CR has a probability $f(\alpha,\Delta r)d\alpha$ to be 
scattered by an angle $\alpha$ to $\alpha+d\alpha$ while propagating a 
distance $\Delta r$. 
CRs moving at an angle $\varphi$ {\it after} a scattering by an
angle
$\alpha$ may have been scattered from a direction $\varphi'$ obeying
$\varphi^{\prime2} = \varphi^2 + \alpha^2 - 2\varphi\alpha\cos p$, where $p$
is the angle over a circle of radius $\alpha$ around the direction
$\varphi$. The average over this circle of $\zeta' - \zeta$ is $\alpha^2$,
and the average of $(\zeta' - \zeta)^2$ is $2\zeta\alpha^2$ (where we
use the approximation $\alpha \ll \varphi$, valid for our case of many
scatterings along the CR trajectory). Therefore, the difference
between the number of CRs scattered into $\varphi$ and the
number scattered out of this direction over the interval $\Delta r$ is
$<\!\alpha^2(\Delta r)\!> (\partial P / \partial \zeta +
\zeta \partial^2 P/\partial \zeta^2)$, where $<\!\alpha^2(\Delta r)\!>$ denotes
the average of $\alpha^2$ over the distribution $f(\alpha,\Delta r)$.
Combining this expression with (\ref{dPp}), our final equation is
\begin{equation}
{\partial P\over \partial r} = {2P\over r} +
\left( {2\zeta\over r} + \psi \right)\,
{\partial P\over \partial \zeta} +
\zeta\psi \,
{\partial^2 P\over \partial \zeta^2} -
{\zeta\over 2c} {\partial P \over \partial t}\quad,
\label{dP}
\end{equation}
where $\psi\equiv<\!\alpha^2(\Delta r)\!>/\Delta r$. 
The effect of the magnetic 
field is therefore to introduce a ``mean scattering
rate,'' $\psi$. If the IGMF is characterized by a
single correlation length $\lambda$, then $\psi=
<\!\alpha^2(\lambda)\!>/\lambda=2(eB/E)^2\lambda/3$, where
the $2/3$ factor accounts for random field orientations. 
If the field fluctuates
on a range of scales $\lambda$, with some power spectrum $dB^2(\lambda)
=\Phi_\lambda d\lambda$,
then  
\begin{equation}
\psi\sim {2e^2\over3E^2}\int d\log\lambda\  \Phi_\lambda\lambda^2
\equiv {2e^2\over3E^2}\overline{B^2\lambda}\quad.
\label{scat}
\end{equation}

Equation (\ref{dP}) is analogous to the equation describing the propagation of 
photons in a scattering medium in the limit of infinite optical depth, and
was derived for the latter case by Alcock \& Hatchett (1978), who presented
an analytical solution (their eqs. [21]-[24]).
Of particular interest to us is the
integration of $P$ over $\zeta$, giving the distribution of time delays.
We can express this in terms of the energy distribution of CRs
with fixed time delay, which is the most relevant observable quantity
given the long time delays expected in the bursting model. The differential
flux received from a burst, which produced $N(E)dE$ CRs in the energy range
$E$ to $E+dE$, is:
\begin{equation}
F(E; D, t)= {3\pi c \over2e^2\overline{B^2\lambda}D^4}E^2 N(E)
\, \sum\limits_{n=1}^{\infty} (-1)^{n+1}\, n^2\,
\exp\left[ -2n^2\pi^2{E^2\over E_0^2(t,D)} \right]\quad,
\label{flux}
\end{equation}
where $E_0(t,D)=De(2\overline{B^2\lambda}/3ct)^{1/2}$. 
Current observations require the
generation spectrum $N(E)$ to satisfy $N(E)\propto E^{-n}$ with $n\simeq2$
(\cite{Wb}). Therefore, $E^2N(E)$ on the right hand side of (\ref{flux})
is approximately constant, and the differential spectrum of the source is given
by $F(E)/E^2N(E)$, presented in Figure 1. For this spectrum, the ratio of the 
rms CR energy spread to the average energy is $30\%$.

  The form of the angular image of CRs is also predicted.
The angular distribution of CRs of a fixed energy averaged over all time
delays is a gaussian, $P(\zeta;E)\propto\exp[-\zeta/\theta_s^2(E)]$ with 
$\theta_s(E)=e(2D\overline{B^2\lambda}/9)^{1/2}/E$.
The angular distribution of CRs at a fixed time delay averaged over all
energies, which is more interesting given the long time delays 
expected, is approximately given by $P(\zeta;t)\propto\exp\{-0.58[\zeta/
\theta_s^2(\bar E)]^2\}$, where $\bar E$ is the average energy of CRs observed
at a fixed time delay $t$. The joint energy and angular distribution of
CRs at a fixed time delay, which may be observable, can also be predicted
in detail from the solution of Alcock \& Hatchett (1978).

\section{Discussion}

We have shown that in a bursting source model 
the spectral shape and angular size of bright CR sources
below the pion production threshold, $E\lesssim4\times10^{19}{\rm eV}$,
are sensitive to the ratio of the typical displacement of
CRs, $\theta_sD$, and the IGMF correlation length $\lambda$. 
The spectral width $\Delta E$ and angular size $\Delta\theta$
are small, $\Delta E/E\sim\Delta\theta/\theta_s\lesssim1\%$, in the limit
$\theta_sD/\lambda\ll 1$, and large, $\Delta E/E\sim
\Delta\theta/\theta_s\sim1$,
in the opposite limit $\theta_sD/\lambda\gg 1$. The spectral
shape of the sources in the limit 
$\theta_sD/\lambda\gg 1$ is given by eq. (\ref{flux}) and shown
in Figure 1. In this limit, the number of CRs observed at a fixed 
time delay with deflection $\varphi$ to $\varphi+d\varphi$ relative to the
direction to the source is approximately proportional to
$\varphi\exp\{-0.58[\varphi^2/\theta_s^2(\bar E)]^2\}d\varphi$, 
where $\bar E$ is the average energy of CRs observed
at the specified time delay. If magnetic fields were
strong in intermediate scales, $\theta_sD/\lambda\sim 1$,
a single source could have several images with angular separation
$\sim\theta_s$, which should be magnified and distorted. As in the
case of gravitational lensing, an odd number of images should be
expected with highly magnified images appearing in merging pairs near
caustics of the magnetic deflection mapping.
Each image could have a narrow energy spectrum $\Delta E/E\ll1$, but
usually the magnetic field on much smaller scales would be important in
widening each image both in angle and energy.
The average energy of CRs in different
images would differ by $\Delta E\sim E$.

The condition $\theta_s D=1$ may be written as
\begin{equation}
{B\over\sqrt{\lambda}}=10^{-10}\left({D\over30{\rm Mpc}}\right)^{-3/2}
E_{19}\quad {\rm G\, Mpc}^{-1/2}.
\label{cond}
\end{equation}
Thus, measuring the spectral width of bright CR sources would allow
to constrain the IGMF parameters provided the
distance scale to the bright sources is estimated.
In a bursting source model,
the typical distance to the sources is proportional to
$(\nu\tau_0)^{-1/5}$, 
where $\nu$ is the bursting rate per unit volume
and $\tau_0\equiv\tau(E=10^{20}{\rm eV}, D=100{\rm Mpc})$
the characteristic time delay 
(cf. eq. [\ref{Dm}]). Current
observations require $\nu\tau_0>10^{-6}{\rm Mpc}$, which sets
an upper limit to the distance of the brightest sources,
$D_m(E)\lesssim30E_{19}^{2/5}{\rm Mpc}$. Future
CR experiments would allow to obtain better constraints on $\nu\tau_0$
by analyzing the number of sources at different energies
(\cite{MW}). Furthermore, if bursts originate in galaxies (as it seems
likely if they are GRBs), their parent galaxies should be identified
given that they are nearby, and the distance to each burst would then
be known. For the GRB
model, where $\nu\simeq3\times10^{-8}{\rm Mpc}^{-3}{\rm yr}^{-1}$,
the upper limit
on $B\lambda^{1/2}$ from Faraday rotation measures, $B\lambda^{1/2}
\le10^{-9}{\rm G\ Mpc}^{1/2}$ (\cite{Kron}, \cite{Vallee}),
sets an upper limit $\tau_0\le10^5{\rm yr}$ to the time delay and
a lower limit to the distance scale $D_m(E)\gtrsim10E_{19}^{2/5}{\rm Mpc}$.
Figure 2 shows the line $\theta_sD=\lambda$ in the $B-\lambda$ plane
together with the Faraday rotation upper limit $B\lambda^{1/2}
\le10^{-9}{\rm G\ Mpc}^{1/2}$ and the lower limit $B\lambda^{1/2}
\ge10^{-11}{\rm G\ Mpc}^{1/2}$ required in the GRB model to allow
$\tau_0\ge50{\rm yr}$. The $\theta_sD=\lambda$ line divides the
allowed region in the $B-\lambda$ plane at $\lambda\sim1{\rm Mpc}$.
Thus, measuring the spectral width of bright CR sources would allow to 
determine
if the field correlation length is much larger, much smaller, or comparable
to $1{\rm Mpc}$.

The proposed Auger experiment
(\cite{huge,huge2})
should be able to detect many CRs from
individual sources at energies of a few times $10^{19}$ eV. In fact,
the number of CRs produced by the brightest source at energy $E$ and
detected by a detector with area $A$ and exposure time $T$ is
approximately $N_m(E)\sim fATN(>E)/4\pi D_m(E)^2\tau(D_m,E)$, where
$N(>E)$ is the number of CRs produced by the source above $E$ and $f$ the
fraction of time during which the source is within the field of view. 
Using eq. (\ref{Dm}) and $\nu N(>E)\simeq5\times10^{36}E_{19}^{-1}
{\rm Mpc}^{-3}{\rm yr}^{-1}$,
as required by current observations (\cite{Wb}), we find
$N_m(E)\sim10^{3}E_{19}^{-3/5}(\nu\tau_0/10^{-6}{\rm Mpc}^{-3})^{-1/5}
(fAT/10^4{\rm km}^2{\rm yr})$. The Auger experiment 
will have $5000{\rm km}^2$ detectors and energy resolution of
$\sim10\%$, giving an expected number of events detected from
individual bright sources $10^2-10^3$ (assuming a burst 
rate comparable to that of cosmological GRBs), which should easily allow
to determine the spectral width of the sources. 
Furthermore, with an expected angular resolution of
$\sim1\arcdeg$, the apparent angular size of the sources may be
measured in the case $\theta_sD\gg\lambda$: the angular size of the
brightest source at energy $E$ is expected to be 
$\sim\theta_s[D_m(E),E]\sim0.\arcdeg7E_{19}^{-4/5}(\nu\tau_0/10^{-6}
{\rm Mpc}^{-3})^{-1/10}(\tau_0/{\rm 50 yr})^{1/2}$. Measuring
the angular size of the sources would allow to break the 
$\nu\tau_0$ ``degeneracy,'' and to estimate separately the characteristic
time delay $\tau_0\propto B\lambda^{1/2}$ and the burst rate $\nu$.

  Finally, we note that the deflections due to the magnetic field in the
Milky Way will be superposed with those of the IGMF. While the additional
time-delay caused by our galaxy is negligible, the deflection is not.
For example, for a $3\mu G$ field over a scale of $1$ Kpc and a CR energy
$2\times 10^{19}$ eV, the deflection is $8\arcdeg$. This means that if
the IGMF is present only on small scales, large coherent deflections could
still be caused by the galactic field, which could produce elongated images
with the CR energy being correlated with the observed deflection from
the true source position, as well as multiple images of individual bursts.

\acknowledgements
This research
was partially supported by a W. M. Keck Foundation grant 
and NSF grant PHY95-13835 to the IAS.

\newpage

\begin{figure}
\plotone{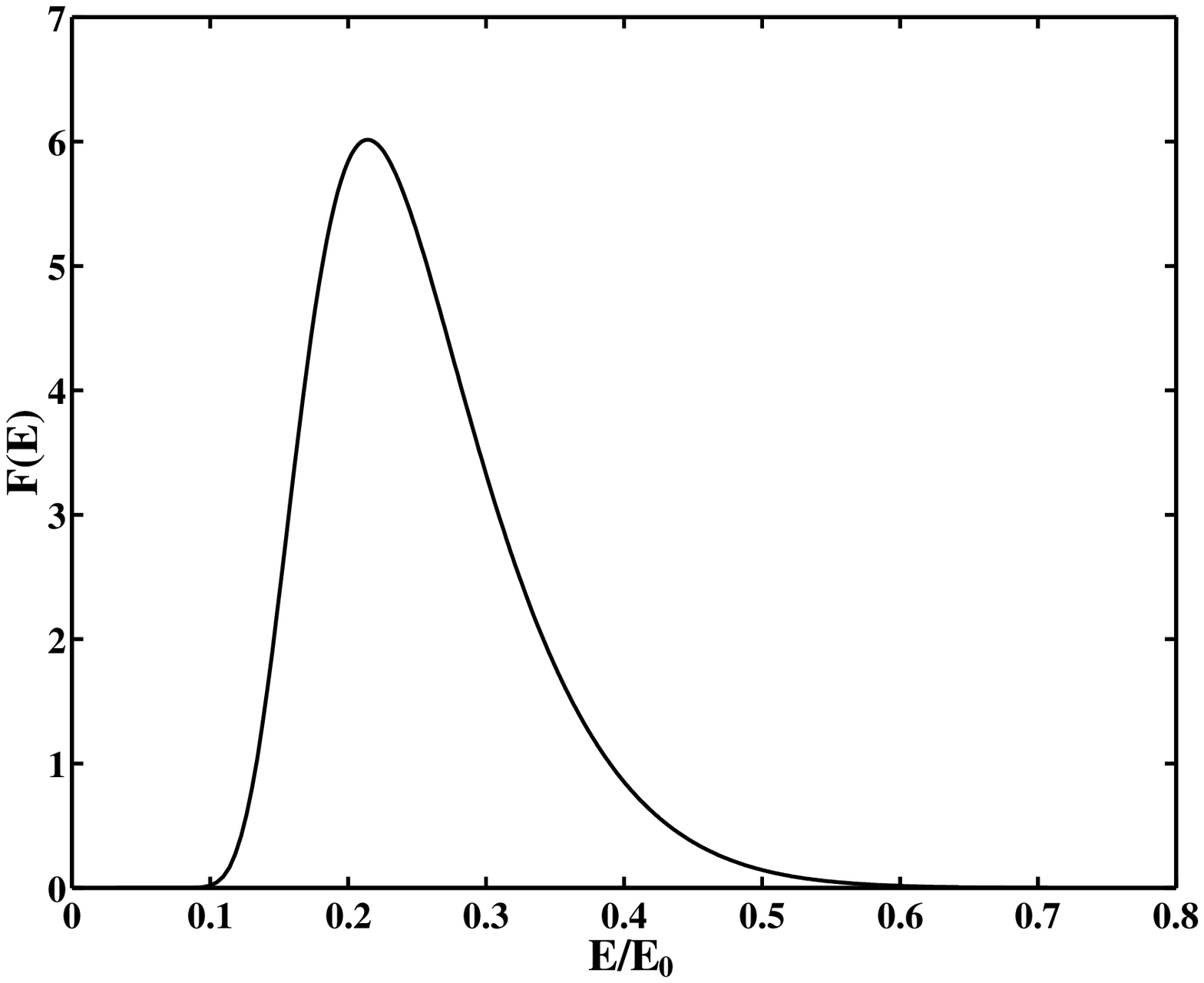}
\caption{
The shape of the differential CR flux received at a fixed time delay $t$ 
from a burst at distance $D$ in the limit $\theta_sD\gg\lambda$; 
$E_0(t,D)\equiv De(2\overline{B^2\lambda}/3ct)^{1/2}$.
}
\label{fig1}
\end{figure}

\begin{figure}
\plotone{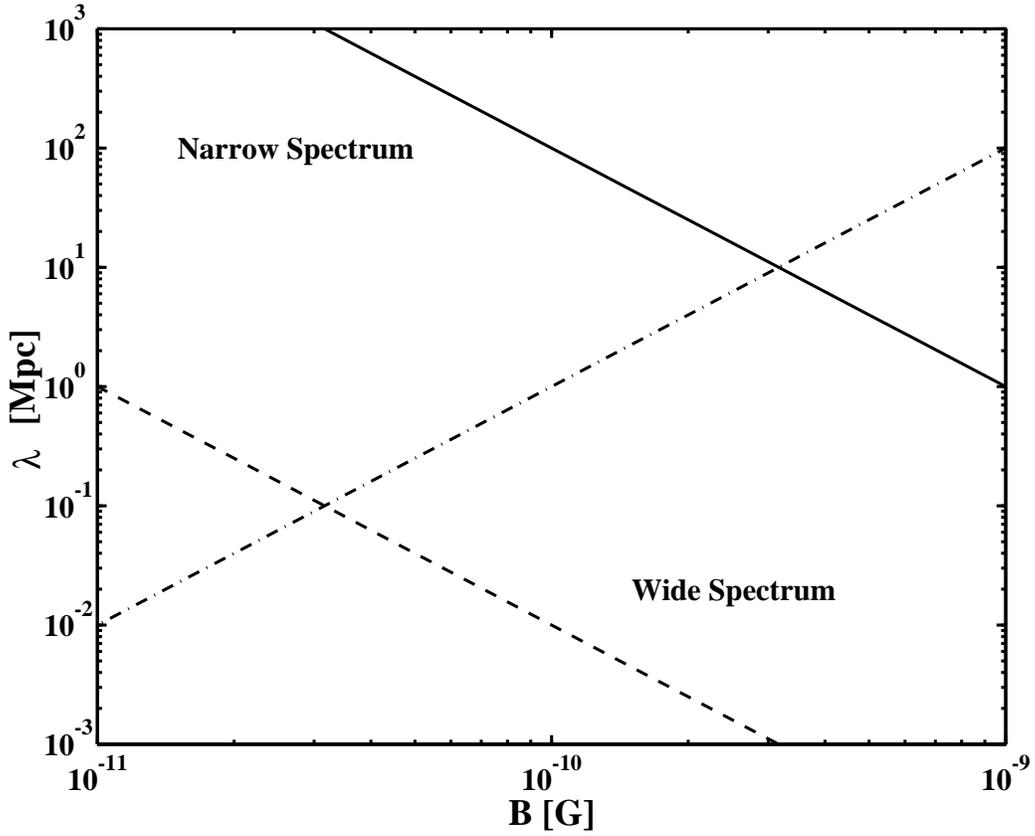}
\caption{
The line $\theta_sD=\lambda$ for a source at a distance $D=30{\rm Mpc}$
observed at energy $E\simeq10^{19}{\rm eV}$ (dot-dash line), shown with
the Faraday rotation upper limit $B\lambda^{1/2}
\le10^{-9}{\rm G\ Mpc}^{1/2}$ (solid line), and with the lower limit 
$B\lambda^{1/2}\ge10^{-11}{\rm G\ Mpc}^{1/2}$ required in the GRB model to 
allow $\tau_0\ge50{\rm yr}$ (dashed line).
}
\label{fig2}
\end{figure}

\end{document}